%% file: main.tex
\documentclass[10pt,letterpaper,twocolumn]{article}
\pdfoutput=1
\usepackage{mathptm}  
\usepackage{usenix}
\usepackage[usenames]{color}
\usepackage{url} 
\usepackage{mdwlist} 
\usepackage{graphicx}
\usepackage{xspace}
\usepackage{tabularx}
\usepackage{multirow}
\usepackage{subfig}
\usepackage{wrapfig}

\usepackage{float}

\floatstyle{ruled}
\newfloat{code}{tbhp}{lop}
\floatname{code}{Code}

\floatstyle{ruled}
\newfloat{logic}{tbhp}{lop}
\floatname{eq}{Logic}

\newcommand{\projname}{\textsc{Quire}\xspace} 

\begin{document}

\date{}

\title{\Large \bf \projname: Lightweight Provenance for Smart Phone
  Operating Systems}

\author{} 
\author{
{\rm Michael Dietz}\\
Rice University
\and
{\rm Shashi Shekhar}\\
Rice University
\and
{\rm Yuliy Pisetsky}\\
Rice University
\and
{\rm Anhei Shu}\\
Rice University
\and
{\rm Dan S.\ Wallach}\\
Rice University
} 

\maketitle

\newcommand{\todo}[1]{\colorbox{yellow}{TODO: #1}}

\begin{abstract}
\input abstract
\end{abstract}

\section{Introduction}
\label{sec:intro}
\input{intro}

\section{Design}
\label{sec:design}
\input{design}

\section{Implementation}
\label{sec:implementation}
\input{implementation}

\section{Applications}
We built two different applications to demonstrate the benefits of
\projname's infrastructure.

\label{sec:applications}
\subsection{Click fraud prevention}
\input{ui}

\subsection{PayBuddy}
\input{paybuddy}

\section{Performance evaluation}
\label{sec:performance}
\input{performance}

\section{Related work}
\label{sec:related}
\input{related}

\section{Future work}
\label{sec:future}
\input{future}

\section{Conclusion}
\label{sec:conclusion}
\input{conclusion}

\bibliographystyle{abbrv}
\bibliography{citations,security}

\newpage


\end{document}

%% file: abstract.tex
Smartphone apps often run with full privileges to access
the network and sensitive local resources, making it difficult for
remote systems to have any trust in the provenance of network
connections they receive.  Even within the phone, different apps with
different privileges can communicate with one another, allowing one
app to trick another into improperly exercising its privileges (a Confused Deputy attack).
In \projname, we engineered two new security mechanisms into Android
to address these issues.
First, we track the call chain of IPCs,
allowing an app the choice of operating with the diminished privileges
of its callers or to act explicitly on its own behalf. Second, a
lightweight signature scheme allows any app to create a signed statement
that can be verified anywhere inside the phone. Both of these
mechanisms are reflected in network RPCs, allowing remote systems
visibility into the state of the phone when an RPC is made.
We demonstrate the usefulness of \projname with two example
applications. We built an advertising service, running distinctly from the app
which wants to display ads, which can validate clicks passed to it
from its host. We also built a payment service, allowing an app to issue a
request which the payment service validates with the user.
An app cannot not forge a payment request by directly connecting to
the remote server, nor can the local payment service tamper with the
request.

%% file: intro.tex
On a smartphone, applications are typically given broad permissions to
make network connections, access local data repositories, and issue
requests to other apps on the device.  
%
%
For Apple's iPhone, the only mechanism that protects users from malicious
apps is the vetting process for an app to get into Apple's app store. (Apple also has the ability
to remotely delete apps, although it's something of an emergency-only system.)
However, 
%
any iPhone app might have its own security vulnerabilities, perhaps through a buffer overflow attack, which can give an attacker full access to the
entire phone.

The Android platform, in contrast, has no significant vetting process before an
app is posted to the Android Market.
Instead, applications from different
authors run with different Unix user ids, containing the damage if an
application is compromised. (In this aspect, Android follows a design similar to SubOS~\cite{ioannidis.bellovin.ea:sub-operating}.)
However, this does nothing to defend a trusted app from
being manipulated by a malicious app via IPC (i.e., 
a Confused Deputy attack~\cite{hardy88}). Likewise, there is no mechanism
to prevent an IPC callee from misrepresenting the intentions of its caller
to a third party.

This mutual distrust arises in many mobile applications.  Consider the
example of a mobile advertisement system.  An application hosting an
ad would rather the ad run in a distinct process, with its own user-id,
so bugs in the ad system do not impact the host. Similarly, the ad system
might not trust its host to display the ad correctly, and must be
concerned with hosts that try to generate fake clicks to inflate their ad revenue.

To address these concerns, we introduce \projname, a low-overhead
security mechanism that provides important context in the form
of {\em provenance} and OS managed data security to local and remote apps
communicating by IPC and RPC respectively. \projname uses two techniques to provide security to communicating
applications.

First, \projname transparently annotates IPCs occurring within the
phone such that the recipient of an IPC request can observe the full
call chain associated with the request. When an application wishes to make a network RPC, it might
well connect to a raw network socket, but it would lack credentials
that we can build into the OS, which can speak to the state of an RPC
in a way that an app cannot forge.
(This contextual information can be thought of as a
generalization of the information provided by the recent HTTP Origin
header~\cite{barth08csrf}, used by web servers to help defeat
cross-site request forgery (CSRF) attacks.)

Second, \projname uses simple cryptographic mechanisms to protect
data moving over IPC and RPC channels.  \projname provides a mechanism for an
app to tag an object with cheap message authentication codes, using keys
that are shared with a trusted OS service.
When data annotated in this manner moves off the
device, the OS can verify the signature and speak to the
integrity of the message in the RPC.

\paragraph{Applications.}
\projname enables a variety of useful applications.  Consider the case
of in-application advertising.  A large number of free applications include
advertisements from services like AdMob.  AdMob is presently implemented as a
library that runs in the same process as the application hosting the ad,
creating trivial opportunities for the application to spoof information to the
server, such as claiming an ad is displayed when it isn't, or claiming
an ad was clicked when it wasn't.  In \projname, the advertisement
service runs as a separate application and interacts with the
displaying app via IPC calls.  The remote application's server can now reliably
distinguish RPC calls coming from its trusted agent, and can
further distinguish legitimate clicks from forgeries,
because every UI event is tagged with a MAC, for which the OS will vouch.

Consider also the case of payment services.  Many smartphone apps would like a
way to sell things, leveraging payment services from PayPal, Google
Checkout, and other such services.  We would like to enable an application to
send a payment request to a local payment agent, who can then pass the
request on to its remote server. The payment agent must 
be concerned with the main app trying to issue fraudulent payment requests,
so it needs to validate requests with the user. Similarly, the main app might
be worried about the payment agent misbehaving, so it wants to create
unforgeable ``purchase orders'' which the payment app cannot corrupt.
All of this can be easily accomplished with our new mechanisms.

\paragraph{Challenges.}
For \projname to be successful, we must accomplish a number of goals.
Our design must be sufficiently general to capture a variety of use
cases for augmented internal and remote communication. Toward that
end, we build on many concepts from Taos~\cite{taos}, including its
compound principals and logic of authentication (see
Section~\ref{sec:design}).
Our implementation must be fast. Every IPC call in the
system must be annotated and must be subsequently verifiable without having a significant
impact on throughput, latency, or battery
life. (Section~\ref{sec:implementation} describes \projname's implementation,
and Section~\ref{sec:performance} presents our performance
measurements.)
\projname expands on related work from a variety of fields, including
existing Android research,
web security, distributed authentication logics, and trusted platform
measurements (see Section~\ref{sec:related}).
%
We expect \projname to serve as a platform for future work in secure
UI design, as a substrate for future research in web browser
engineering, and as starting point for a variety of applications (see
Section~\ref{sec:future}).

%% file: design.tex
Fundamentally, the design goal of \projname is to allow apps to reason
about the call-chain and data provenance of requests,
occurring on the host platform via IPC or on a remote server via RPC,
before committing to a security-relevant decision.  This design goal is shared by a variety of other systems,
ranging from Java's stack inspection~\cite{wallach98,wallach-tosem2000} to many newer systems
that rely on data tainting or information flow control~ (see, e.g.,~\cite{jflow,liskov-flowcontrol,taintdroid}).
In \projname, much like in stack inspection, we wish to support legacy
code without much, if any modification. However, unlike stack
inspection, we don't want to modify the system to annotate and track
every method invocation, nor would we like to suffer the
runtime costs of dynamic data tainting 
as in TaintDroid~\cite{taintdroid}. Likewise, we wish to operate correctly
with apps that have natively compiled code, not just Java code (an issue with
traditional stack inspection and with TaintDroid). Instead, we observe that we
only need to track calls across IPC boundaries,
which happen far less frequently
than method invocations, and which already must pay significant
overheads for data marshaling, context switching, and copying.

Stack inspection has the property that the available privileges at the
end of a call chain represent the intersection of the privileges of every
app along the chain (more on this in Section~\ref{sec:IPCprov}), which
is good for preventing Confused Deputy attacks, but doesn't solve a variety
of other problems, such as validating the integrity of individual data
items as they are passed from one app to another or over the network.
For that, we need semantics akin to digital signatures, but we need
to be much more efficient (more on this in Section~\ref{sec:verifiableStatements}).

\paragraph{Versus information flow}
Our design is necessarily less precise than
dynamic taint analysis, but it's also incredibly flexible. We can
avoid the need to annotate code with static security policies, as
would be required in information flow-typed systems like
Jif~\cite{myers98}. We similarly do not need to poly-instantiate
services to ensure that each instance only handles a single security
label as in systems like DStar/HiStar~\cite{dstar2008}.
Instead, in
\projname, an application which handles requests from multiple callers
will pass along an object annotated with the originator's context when it makes downstream
requests on behalf of the original caller.

Likewise, where a dynamic tainting system like TaintDroid~\cite{taintdroid}
would generally allow a sensitive operation, like learning the phone's precise GPS location,
to occur, but would forbid it from flowing to an unprivileged app; \projname will
carry the unprivileged context through to the point where the dangerous operation is
about to happen, and will then forbid the operation. An information flow approach is
thus more likely to catch corner cases (e.g., where an app caches location data,
so no privileged call is ever performed), but is also more likely to have false positives
(where it must conservatively err on the side of flagging a flow that is actually just fine).
A programmer in an information flow system would need to tag these false positive corner cases
as acceptable, whereas a programmer in \projname would need to add additional security
checks to corner cases that would otherwise be allowed.

%
\subsection{Authentication logic and cryptography}
In order to reason about the semantics of \projname, we need a formal
model to express what the various operations in \projname will do.
Toward that end, we use the Abadi et al.~\cite{ablp91}
(hereafter ``ABLP'') logic of authentication, as used in
Taos~\cite{taos}. In this logic, {\em principals} make {\em
  statements}, which can include various forms of quotation (``Alice
{\bf says} that Bob {\bf says} $X$'') and authorization (e.g., ``Alice {\bf says}  that
Bob speaks for Alice'').  ABLP nicely models the behavior of
cryptographic operations, where cryptographic keys speak for other
principals, and we can use this model to reason about cross-process
communication on a device or over the network.

For the remainder of the current section, we will flesh out
\projname's IPC and RPC design in terms of ABLP and the cryptographic
mechanisms we have adopted.

\subsection{IPC provenance}
\label{sec:IPCprov}

The goal of \projname's IPC provenance system is to allow endpoints
that protect sensitive resources, like a user's fine grained GPS data
or contact information, to reason about the complete IPC call-chain of
a request for the resource before granting access to it.

\projname realizes this goal by modifying the Android IPC middle-ware
layer to automatically build calling context as an IPC call-chain is
formed.  Consider a call-chain where three principals $A$, $B$, and
$C$, are communicating.  If $A$ calls $B$ who then calls $C$ without
keeping track of the call-stack, $C$ only knows that $B$ initiated a
request to it, not that the call from $A$ prompted $B$ to make the
call to $C$.  This loss of context can have significant security
implications in a system like Android where permissions are directly
linked to the identity of the principal requesting access to a
sensitive resource.

\begin{figure}
\includegraphics[width=3in]{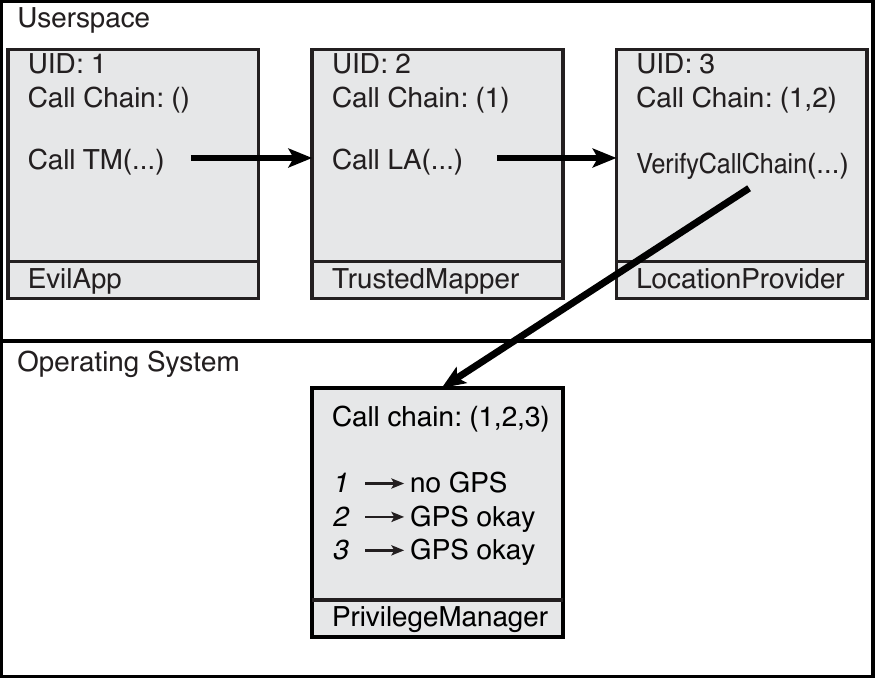}
\caption{Defeating Confused Deputy attacks.\label{fig:confused-deputy}}
\end{figure}

To address this, \projname's design is for any given callee to retain
its caller's call-chain and pass this to a downstream callee. The
downstream callee will automatically have its caller's principal prepended to the ABLP statement.  In our above
scenario, $C$ will receive a statement ``$B$ {\bf says} $A$ {\bf says} {\bf Ok}'',
where {\bf Ok} is an abstract token representing that the given resource
is authorized to be used. It's now the burden of $C$ (or \projname's privilege manager,
operating on $C$'s behalf) to prove {\bf Ok}. As Wallach et al.~\cite{wallach-tosem2000}
demonstrated, this is equivalent to validating that each principal in the calling chain
is individually allowed to perform the action in question.

\paragraph{Confused Deputy} With this additional context, \projname defeats Confused Deputy attacks; if any one of the
principals in the call chain is not privileged for the action being
taken, permission is denied. Figure~\ref{fig:confused-deputy} shows this in the context of
an evil application, lacking fine-grained location privileges, which
is trying to abuse the privileges of a trusted mapping program, which
happens to have that privilege. The mapping application, never
realizing that its helpful API might be a security vulnerability,
na\"{\i}vely and automatically passes along the call chain along to
the location service. The location service then uses the call chain to prove (or
disprove) that the request for fine-grained location show be allowed.

As with traditional stack inspection, there will be times that an app
genuinely wishes to exercise a privilege, regardless of its caller's
lack of the same privilege. Stack inspection solves this with an {\em
enablePrivilege} primitive that, in the ABLP logic, simply doesn't
pass along the caller's call stack information. The callee after
privileges are enabled gets only the caller's identity. (In the
example of Figure~\ref{fig:confused-deputy}, the trusted mapper would
drop evil app from the call chain, and the location service would only
hear that the map application wishes to use the service.)

Our design is, in effect, an example of the ``security passing style''
transformation~\cite{wallach-tosem2000}, where security beliefs are
passed explicitly as an IPC argument rather than passed implicitly as
annotations on the call stack. One beneficial consequence of this is
that a callee might well save the statement made by its caller and
reuse them at a later time, perhaps if they queue requests for later
processing, in order to properly modulate the privilege level of
outgoing requests.

\paragraph{Security analysis}
While apps, by default, will pass along call chain information without
modification, \projname allows a caller to forge the identities of its
upstream callers. No cryptography needs to be used to prevent this. By enabling a
caller to misrepresent its antecedent call chain, this would
seem to be a serious security vulnerability, but 
there is no {\em incentive} for a caller to lie, since
nothing it quotes from its antecedent callers can increase its
privileges in any way.

Conversely, our design requires the callee to learn the caller's
identity in an unforgeable fashion. When the callee prepends the
``Caller {\bf says}'' tokens to the statement it hears from the
caller, using information that is available as part of every Android
Binder IPC, any lack of privileges on the caller's part will be
properly reflected when the privileges for the trusted operation are
later evaluated.

Furthermore, our design is incredibly lightweight; we can construct
and propagate IPC call chains with very little impact on the overall
IPC performance (see Section~\ref{sec:performance}).

\subsection{Verifiable Statements}
\label{sec:verifiableStatements}
Stack inspection semantics are helpful, but are not sufficient for
many security needs. We envision a variety of scenarios where we
will need semantics equivalent to digital signatures, but with much
better performance than public-key cryptographic operations.

\paragraph{Definition}
A {\em verifiable statement} is a 3-tuple $[P, M, A(M)_P]$
where $P$ is the principal that said message $M$, and $A(M)_P$ is an
authentication token that can be used by the Authority Manager OS service to
verify $P$ said $M$. In ABLP, this tuple represents the statement ``$P$ {\bf says} $M$.''

In order to operate without requiring slow public-key cryptographic
operations, we must instead use message authentication codes (MAC).
MAC functions, like HMAC-SHA1, run several orders of magnitude faster
than digital signature functions like DSA, but MAC functions require a
shared key between the generator and verifier of a MAC. To avoid an
$N^2$ key explosion, we instead have every application share a key
with a central, trusted authority manager. As such, any app can
produce a statement ``App {\bf says} $M$'', purely by computing a MAC
with its secret key. However, for a second app to verify it, it must
send the statement to the authority manager. If the authority manager
says the MAC is valid, then the second app will believe the veracity
of the statement.

\subsection{RPC attestations}
\label{sec:RPCatte}
When moving from on-device IPCs to Internet RPCs, some of the properties that
exist on the device disappear. Most notably, the receiver of a call
can no longer open a channel to talk to the authority manager, even if
they did trust it\footnote{Like it or not, with NATs, firewalls, and
  other such impediments to bi-directional connectivity, we can only
  assume that the phone can make outbound TCP connections, not receive
  inbound ones.}. To combat this, \projname's design requires an
additional ``network provider'' system service, which can speak
over the network, on behalf of statements made on the phone.
This will require it to speak with a cryptographic secret that
is not available to any applications on the system.

One method for getting such a secret key is to have the phone
manufacturer embed an X.509 certificate which they sign along with the
corresponding private key into storage which is only accessible to the
OS kernel. This certificate can be used to establish a
client-authenticated TLS connection to a remote service, with the
remote server using the presence of the client certificate, as
endorsed by a trusted certification authority, to provide
confidence that it is really communicating with the \projname phone's
operating system, rather than an application attempting to impersonate
the OS. With this attestation-carrying encrypted channel in place,
RPCs can then carry a serialized form of the same statements passed
along in \projname IPCs, including both call chains and signed statements,
with the network provider trusted to speak on behalf of the activity inside the phone.

All of this can be transmitted in a variety of ways, such as a new HTTP
header.  Regular \projname applications would be able to speak through
this channel, but the new HTTP headers, with their security-relevant
contextual information, would not be accessible to or forgeable by the
applications making RPCs.  (This is analogous to the HTTP origin
header~\cite{barth08csrf}, generated by modern web browsers, but
carries more detailed contextual information from the caller.)

The strength of this security context information
is limited by the ability of the device and the OS to protect the
key material. If a malicious application can extract the private key, then it
would be able to send messages with arbitrary claims about the
provenance of the request. This leads us inevitably to techniques from
the field of trusted platform measurement (TPM), where stored
cryptographic key material is rendered unavailable unless the kernel
was properly validated when it booted. TPM chips are common in many of
today's laptops and could well be installed in future smartphones.

Even without TPM hardware, Android phones generally prohibit
applications from running with full root privileges, allowing the kernel to
protect its data from malicious apps. This is a sound design until users forcibly
``root'' their phones, which is commonly done to work around
carrier-instituted restrictions such as forbidding phones from freely
relaying cellular data services as WiFi hotspots. Regardless, {\em
most} users will never ``root'' their phones, preventing normal
applications, even if they want superuser privileges, from getting
them, and then compromising the network provider's private keys.

\paragraph{Privacy.}
An interesting concern arises with our design: Every RPC call made
from \projname uses the unique public key assigned to that phone.
Presumably, the public key certificate would contain a variety of
identifying information, thus making {\em every} RPC personally
identify the owner of the phone. This may well be desirable in {\em
  some} circumstances, notably allowing web services with Android
applications acting as frontends to completely eliminate any need for
username/password dialogs. However, it's clearly undesirable in other
cases. To address this very issue, the Trusted Computing Group has
designed what it calls ``direct anonymous
attestation''\footnote{\url{http://www.zurich.ibm.com/security/daa/}},
using cryptographic group signatures to allow the caller to prove that
it knows one of a large group of related private keys without saying
anything about which one. A production implementation of \projname
could certainly switch from TLS client-auth to some form of anonymous
attestation without a significant performance impact.

An interesting challenge, for future work, is being able to switch
from anonymous attestation, in the default case, to classical
client-authentication, in cases where it might be desirable. One
notable challenge of this would be working around users who will click
affirmatively on any ``okay / cancel'' dialog that's presented to them
without ever bothering to read it. Perhaps this could be finessed with
an Android privilege that is requested at the time an application is installed.
Unprivileged apps can only make anonymous attestations, while more trusted
apps can make attestations that uniquely identify the specific phone.


%% file: implementation.tex
\projname is implemented as a set of extensions to the existing Android Java
runtime libraries and Binder IPC system.  The authority manager and
network provider are trusted components and therefore implemented as
OS level services while our modified Android interface definition
language code generator provides IPC stub code that allows
applications to propagate and adopt an IPC
call-stack.  The result, which is implemented in
around 1300 lines of Java and C++ code, is an extension to
the existing Android OS that provides locally verifiable statements, IPC provenance, and
authenticated RPC for \projname-aware applications and backward
compatibility for existing Android applications.

\subsection{On- and off-phone principals}
The Android architecture sandboxes
applications such that apps from different sources run as different Unix users.
Standard Android features also allow us to resolve user-ids into human-readable
names and permission sets, based on the applications' origins.  Based
on these features, the prototype \projname implementation defines principals as the tuple of a user-id and process-id.
We include the process-id component to allow the
recipient of an IPC method call to stipulate policies that force the
process-id of a communication partner to remain unchanged across a series of
calls. (This feature is largely ignored in the applications we
have implemented for testing and evaluation purposes, but it might be useful later.)

While principals defined by user-id/process-id tuples are sufficient for the identification
of an application on the phone, they are meaningless to a remote
service.  \projname therefore resolves the user-id/process-id tuples
used in IPC call-chains into an externally meaningful string
consisting of the marshaled chain of application
names when RPC communication is invoked to move data off the phone.
This lazy resolution of IPC principals
allows \projname to reduce the memory
footprint of statements when performing IPC calls at the cost
of extra effort when RPCs are performed.

\subsection{Authority management}
The Authority Manager discussed in Section~\ref{sec:design} is
implemented as a system service that runs within the operating
system's reserved user-id space.  The interface exposed by the service
allows userspace applications to request a shared secret, submit a
statement for verification, or request the resolution of the principal
included in a statement into an externally meaningful form.

When an application requests a key from the authority
manager, the Authority Manager maintains a table mapping user-id / process-id tuples to the key.
It is important to note that a
subsequent request from the same application 
will prompt the Authority
Manager to create a new key for the calling application and replace
the previous stored key in the lookup table.  This prevents attacks
that might try to exploit the reuse of user-ids and process-ids
as applications come and go over time.

\subsection{Verifiable statements}
Section~\ref{sec:design} introduced the idea of attaching an OS
verifiable statement to an object in order to allow principals later in a
call-chain to verify the authenticity and integrity of a received
object.

Our implementation of this abstract concept involves a parcelable
statement object that consists of a principal identifier as well as an
authentication token.  When
this statement object is attached to a parcelable object, the
annotated object contains all the information necessary for the
Authority Manager service to validate the authentication token
contained within the statement.
Therefore the annotated object can be sent over Android's IPC channels and later
delivered to the \projname Authority Manger for verification by the OS
as discussed in section~\ref{sec:design}.  

\projname's verifiable statement implementation establishes the
authenticity of message with a hashed message
authentication code (HMAC) digest rather than a heavyweight
public key digital signature.  This implementation
decision drastically reduces the cost of 
creating and verifying a statement, as discussed in section \ref{sec:performance} while still providing the
authentication and integrity semantics required by \projname. 

\paragraph{Fast authenticator creation}
A fundamental assumption of our decision to use Hashed Message
Authentication Codes (HMACs) rather than public-key digital signatures as our
cryptographic mechanism for authentication was that the
Android-provided HMAC library code would yield results within a
constant factor of OpenSSL's baseline numbers. In practice, doing HMAC-SHA1
in pure Java was still slow enough to be an issue.

We resolved the issue by using the native C implementation from OpenSSL
and exposing it to Java code as a Dalvik VM intrinsic function, rather than a JNI native method.
This eliminated unnecessary copying and runs at full native speed (see Section~\ref{sec:perf-statement}).

\subsection{Code generator}
\label{sec:impl:code-generator}

The key to the stack inspection semantics that \projname provides is
an extension to the Android Interface Definition Language (AIDL) code
generator.  This piece of software is responsible for taking in a
generalized interface definition and creating stub and proxy code to
facilitate Binder IPC communication over the interface as defined in the AIDL
file.

The \projname code generator differs from the stock Android code
generator in that it adds directives to the marshaling and
unmarshaling phase of the stubs that pulls the call-chain context from
the calling app and attaches it to the outgoing IPC message for the
callee to retrieve.  These directives allow for the ``quoting''
semantics that form the basis of a stack inspection based policy system.

Our prototype implementation of the \projname AIDL code generator
requires that an application developer specify that an AIDL method
become ``\projname aware'' by defining the method with a reserved
{\em auth} flag in the AIDL input file.  This flag informs the
\projname code generator to produce additional proxy and stub code
for the given method that enables the propagation and delivery of the
call-chain context to the specified method. A production implementation
would pass this information implicitly on all IPC calls.

%% file: ui.tex

\newcommand{\appframe}{\textit{AppFrame}\xspace}  
\newcommand{\motionevent}{\textit{MotionEvent}\xspace}  
\newcommand{\transtheme}{\textit{Theme.Translucent}\xspace}  
\newcommand{\androidmanifest}{\textit{AndroidManifest.xml}\xspace}  
\newcommand{\gingerbread}{Android 2.3\xspace}  

Current Android-based advertising systems, such as AdMob, are deployed as a library that an app includes as part of its distribution. So far as the Android OS is concerned, the app and its ads are operating within single domain, indistinguishable from one another. Furthermore, because advertisement services need to report their activity to a network service, any ad-supported app must request network privileges, even if the app, by itself, doesn't need them.

From a security perspective, mashing these two distinct security domains together into a single app creates a variety of problems. In addition to requiring network-access privileges, the lack of isolation between the advertisement code and its host creates all kinds of opportunities for fraud. The hosting app might modify the advertisement library to generate fake clicks and real revenue.

This sort of click fraud is also a serious issue on the web, and it's typically addressed by placing the advertisements within an iframe, creating a separate protection domain and providing some mutual protection. To achieve something similar with \projname, we needed to extend Android's UI layer and leverage \projname's features to authenticate indirect messages, such as UI events, delegated from the parent app to the child advertisement app.

\begin{figure}[t]
    \centering
    \includegraphics[width=3in]{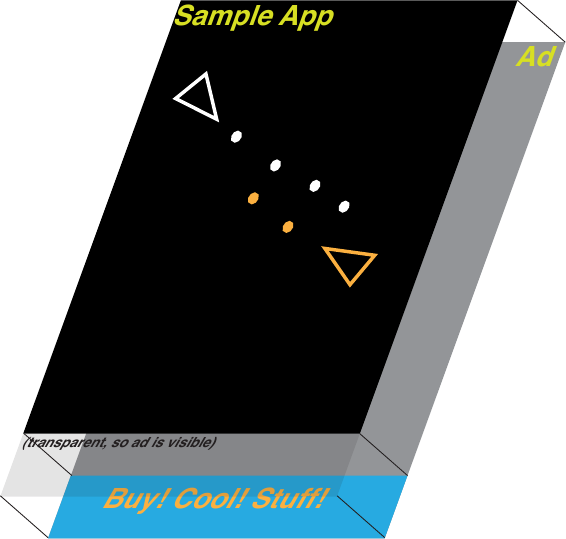}
    \caption{The host and advertisment apps.}
    \label{fig:adverthostlayout}
\end{figure}

\paragraph{Design challenges}
Fundamentally, our design requires two separate apps to be stacked
(see Figure~\ref{fig:adverthostlayout}), with the primary application
on top, and opening a transparent hole through which the subordinate
advertising application can be seen by the user. This immediately raises
two challenges. First, how can the advertising app know that it's
actually visible to the user, versus being obscured by the
application? And second, how can the advertising app know that the
clicks and other UI events it receives were legitimately generated by
the user, versus being synthesized or replayed by the primary application.

\paragraph{Stacking the apps}
This was straightforward to implement. The hosting application
implements a translucent theme (\transtheme), making the background
activity visible. When an activity containing an advertisement is
started or resumed, we modified the activity launch logic system to
ensure that the advertisement activity is placed below the associated
host activities. When a user event is delivered to the \appframe view,
it sends the event along with the current location of \appframe in the
window to the an advertisement event service. This allows our
prototype to correctly display the two apps together.

\paragraph{Visibility}
Android allows an app to continue running, even when it's not on the
screen. Assuming our ad service is built around payments per click,
rather than per view, we're primarily interested in knowing, at the
moment that a click occurred, that the advertisement was actually
visible. \gingerbread added a new feature where motion events contain
an ``obscured'' flag that tells us precisely the necessary
information. The only challenge is knowing that the \motionevent 
we received was legitimate and fresh.

\begin{figure}[t]
    \centering
    \includegraphics[width=3in]{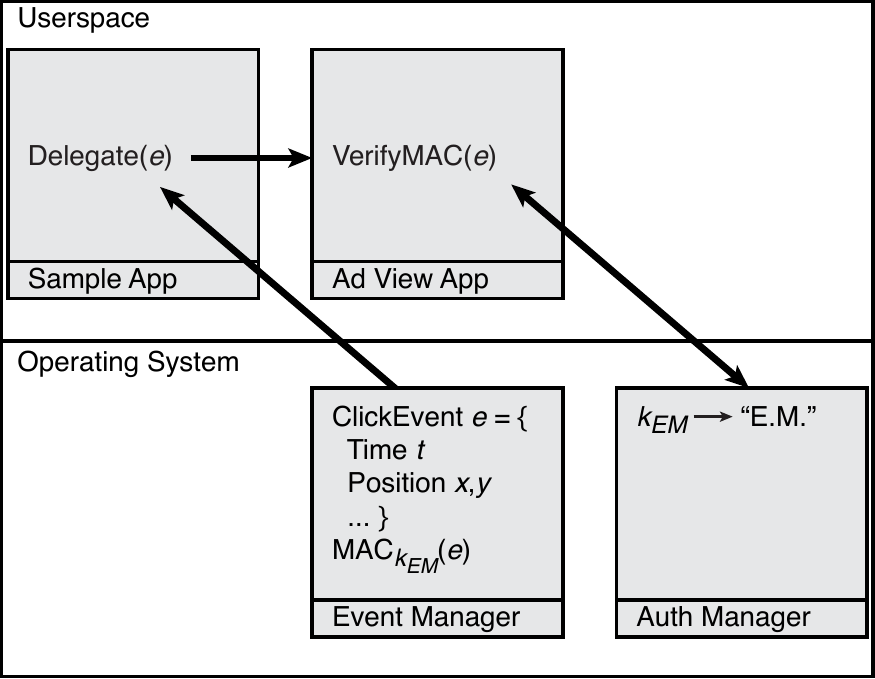}
    \caption{Secure event delivery from host app to advertisement app.}
    \label{fig:secureeventdelivery}
\end{figure}

\paragraph{Verifying events}
With our stacked app design, motion events are delivered to the host
app, on top of the stack. The host app then recognizes when an event
occurs in the advertisement's region and passes the event along. To
complicate matters, \gingerbread reengineered the event system to
lower the latency, a feature desired by game designers. Events are now
transmitted through shared memory buffers, below the Java layer.

In our design, we leverage \projname's signed statements. We modified
the event system to augment every \motionevent (as many as 60 per
second) with one of our MAC-based signatures. This means we don't have
to worry about tampering or other corruption in the event system. Instead,
once an event arrives at the advertisment app, it first validates the
statement, then validates that it's not obscured, and finally validates
the timestamp in the event, to make sure the click is fresh. This process
is summarized in Figure~\ref{fig:secureeventdelivery}.

At this point, the local advertising application can now be satisfied
that the click was legitimate and that the ad was visible when the
click occurred and it can communicate that fact over the Internet,
unspoofably, with \projname's RPC service.

All said and done, we added around 500 lines of Java code for 
modifying the activity launch process, plus a modest amount of C code
to generate the signatures. While our
implementation does not deal with every possible scenario (e.g., changes
in orientation, killing of the advertisement app due to low memory, and other such things)
it still demonstrates the feasibility of hosting of advertisement in
separate processes and defeating click fraud attacks.

%% file: paybuddy.tex
To demonstrate the usefulness of \projname for RPCs, we implemented a
micropayment application called PayBuddy: a standalone Android
application which exposes an activity to other applications on the device
to allow those applications to request payments. By developing this as a
separate application we avoid many types of attacks which circumvent user 
approval of payments. 

\begin{figure}[t]
    \centering
    \includegraphics[width=3in]{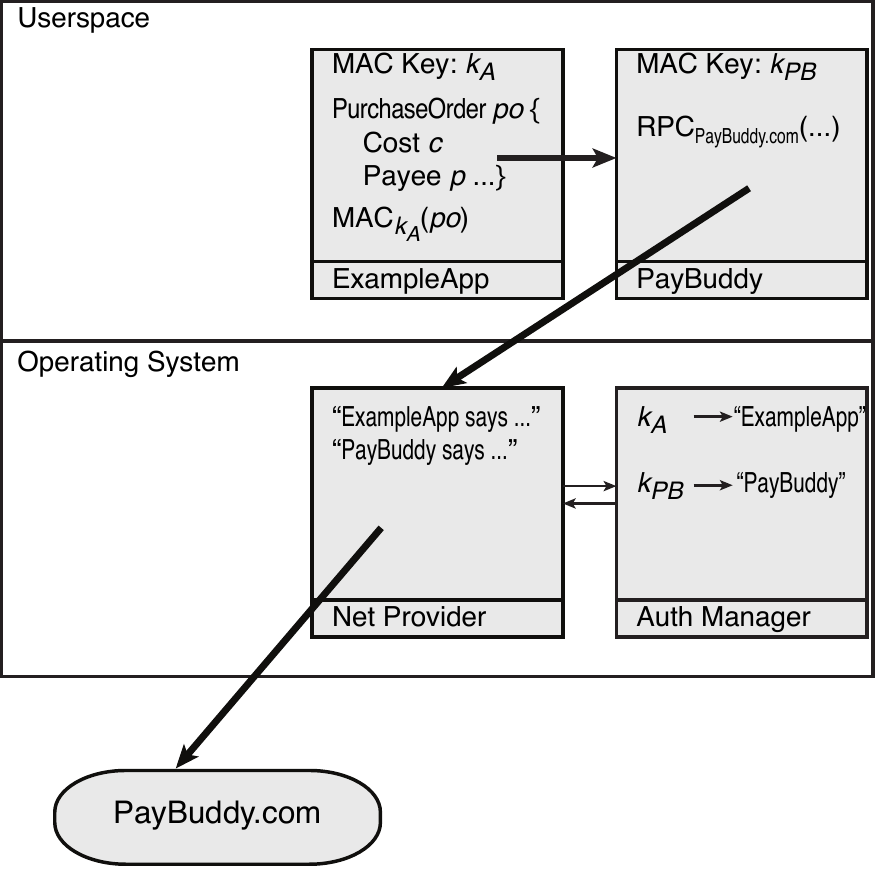}
    \caption{Message flow in the PayBuddy system.}
    \label{fig:paybuddy}
\end{figure}

To demonstrate how PayBuddy works, consider the example shown in
Figure~\ref{fig:paybuddy}. Application ExampleApp wishes to allow the user
to make an in-app purchase. To do this, ExampleApp creates and serializes a
purchase order object and signs it with its MAC key $k_A$. It then sends the
signed object to the PayBuddy application, which can then prompt the user to
confirm their intent to make the payment. After this, PayBuddy passes the
purchase order along to the operating system's Network Provider.
At this point, the Network Provider can verify the signature on the purchase
order, and also that the request came from the PayBuddy application. It then
sends the request to the PayBuddy.com server over a client-authenticated
HTTPS connection. The contents of ExampleApp's purchase order are included in
an HTTP header, as is the call chain (``ExampleApp, PayBuddy'').

At the end of this, PayBuddy.com knows the following:

\begin{itemize}
    \item The request came from a particular device with a given certificate.
    \item The purchase order originated from ExampleApp and was not tampered with by
    the PayBuddy application.
    \item The PayBuddy application approved the request (which means that the
            user gave their explicit consent to the purchase order).
\end{itemize}

At the end of this, if PayBuddy.com accepts the transaction, it can
take whatever action accompanies the successful payment (e.g.,
returning a transaction ID that ExampleApp might send to its home
server in order to download a new level for a game).

\paragraph{Security analysis}
Our design has several curious properties. Most notably, the
ExampleApp and the PayBuddy app are mutually distrusting of each
other.

The PayBuddy app doesn't trust the payment request to be legitimate,
so it can present an ``okay/cancel'' dialog to the user. In that
dialog, it can include the cost as well as the ExampleApp name, which
it received through the \projname call chain. The PayBuddy app will
only communicate with the PayBuddy.com server if the user approves the
transaction.

Similarly, ExampleApp has only a limited amount of trust in the
PayBuddy app. By signing its purchase order, and including a unique
order number of some sort, a compromised PayBuddy app cannot modify or
replay the message. Because the OS's net provider is trusted to speak
on behalf of both the ExampleApp and the PayBuddy app, the remote
PayBuddy.com server gets ample context to understand what happened on
the phone and deal with cases where a user later tries to repudiate a payment.

Lastly, the user's
PayBuddy credentials are never visible to ExampleApp in any way. Once the PayBuddy
app is bound, at install time, to the user's matching account on PayBuddy.com, there
will be no subsequent username/password dialogs. All the user will see is an okay/cancel
dialog. Once users are accustomed to this, they will be more likely to react
with skepticism when presented with a phishing attack that demands their PayBuddy credentials.
(A phishing attack that's completely faithful to the proper PayBuddy user interface would
only present an okay/cancel dialog, which yields no useful information for the attacker.)

%% file: performance.tex
\subsection{Experimental methodology}
All of our experiments were performed on the standard Android developer
phone, the Nexus One\footnote{\url{http://www.google.com/phone/static/en_US-nexusone_tech_specs.html}},
which has a 1GHz ARM core (a Qualcomm QSD 8250), 512MB of RAM, and 512MB of
internal Flash storage.  We conducted our experiments with the phone
displaying the home screen and running the normal set of applications
that spawn at start up.  We replaced the default ``live wallpaper''
with a static image to eliminate its background CPU load.

All of our benchmarks are measured using the Android Open Source
Project's (AOSP) Android 2.3 (``Gingerbread'') as pulled from the AOSP
repository on December 21st, 2010.  \projname is implemented as a
series of patches to this code base.  We used an unmodified Gingerbread build for ``control''
measurements and compared that to a build with our \projname features
enabled for ``experimental'' measurements.

\subsection{Microbenchmarks}

\subsubsection{Signed statements}
\label{sec:perf-statement}
Our first micro benchmark of \projname measures the cost of creating
and verifying statements of varying sizes. To do this, we had an application
generate random byte arrays of varying sizes from 10 bytes to 8000 bytes and
measured the time to create 1000 signatures of the data, followed by
1000 verifications of the signature. Each set of measured signatures and
verifications was preceded by a priming run to remove any first-run effects.
We then took an average of the middle 8
out of 10 such runs for each size. The large number of runs is due to variance
introduced by garbage collection within the Authority Manager. Even with this
large number of runs, we could not fully account for this, leading to some
jitter in the measured performance of statement verification. 

\begin{figure}[t]
    \centering
    \includegraphics[width=3in,height=2.25in]{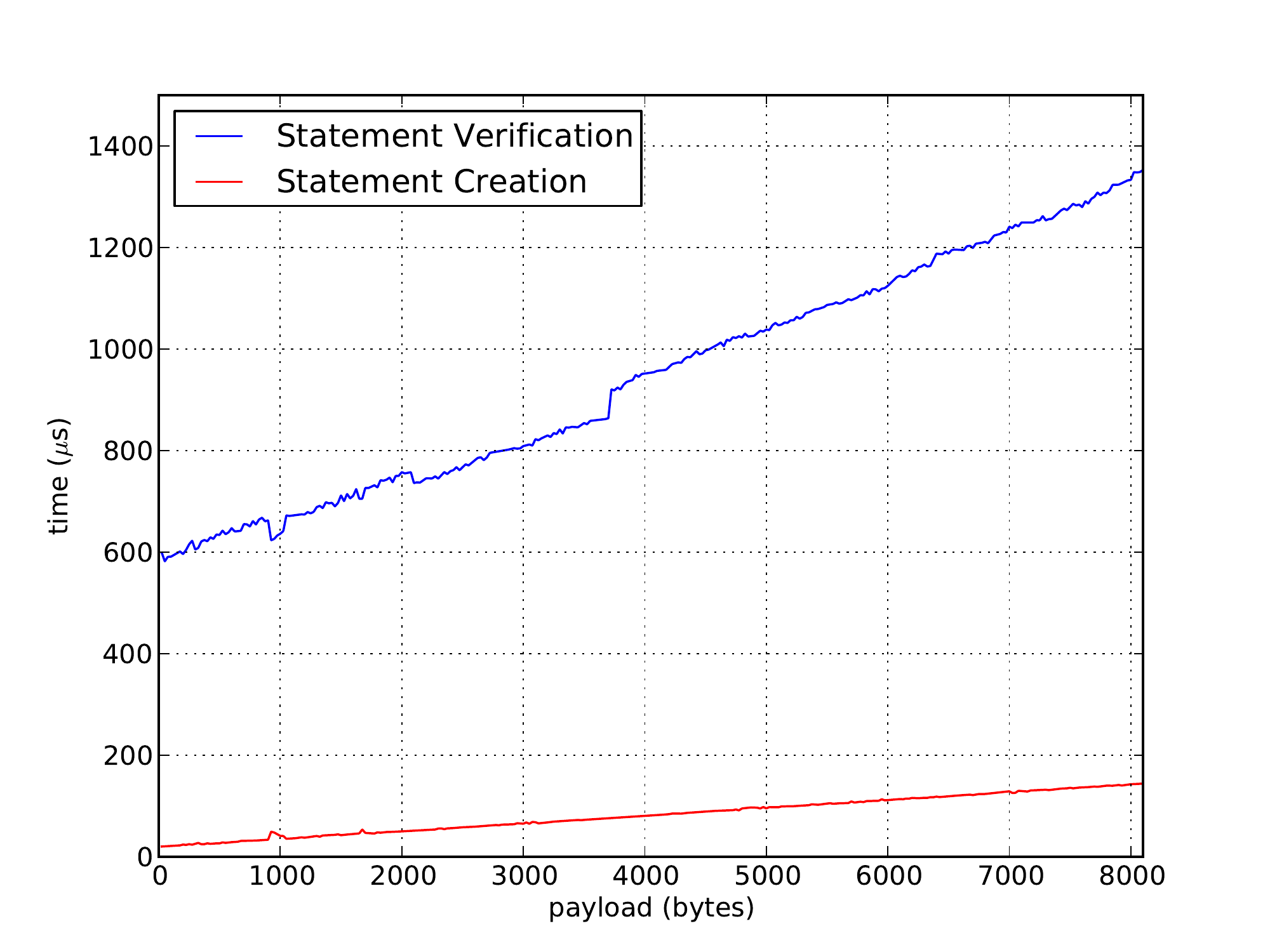}
    \caption{Statement creation and verification time vs payload size.}
    \label{fig:perf-statements}
\end{figure}

The results in Figure \ref{fig:perf-statements} show that statement creation
carries a minimal fixed overhead of 20 microseconds with an additional cost
of 15 microseconds per kilobyte. Statement verification, on the other hand,
has a much higher cost: 556 microseconds fixed and an additional 96
microseconds per kilobyte. This larger cost is primarily due to the context switch
and attendant copying overhead required to ask the Authority Manager to perform the verification. However, with statement verification being a much
less frequent occurrence than statement generation, these performance numbers
are well within our performance targets.

\subsubsection{IPC call-chain tracking}
\label{sec:perf-auth-ipc}
Our next micro-benchmark measures the additional cost of tracking the call
chain for an IPC that otherwise performs no computation. We implemented a
service with a pair of methods, of which one uses the \projname IPC extensions
and one does not. These methods both allow us to pass a byte array of arbitrary
size to them. We then measured the total round trip time needed to make each
of these calls.
These results are intended to demonstrate the slowdown introduced by the
\projname IPC extensions in the worst case of a round trip null
operation that takes no action on the receiving end of the IPC method
call.

\begin{figure}[t]
    \centering
    \includegraphics[width=3in,height=2.25in]{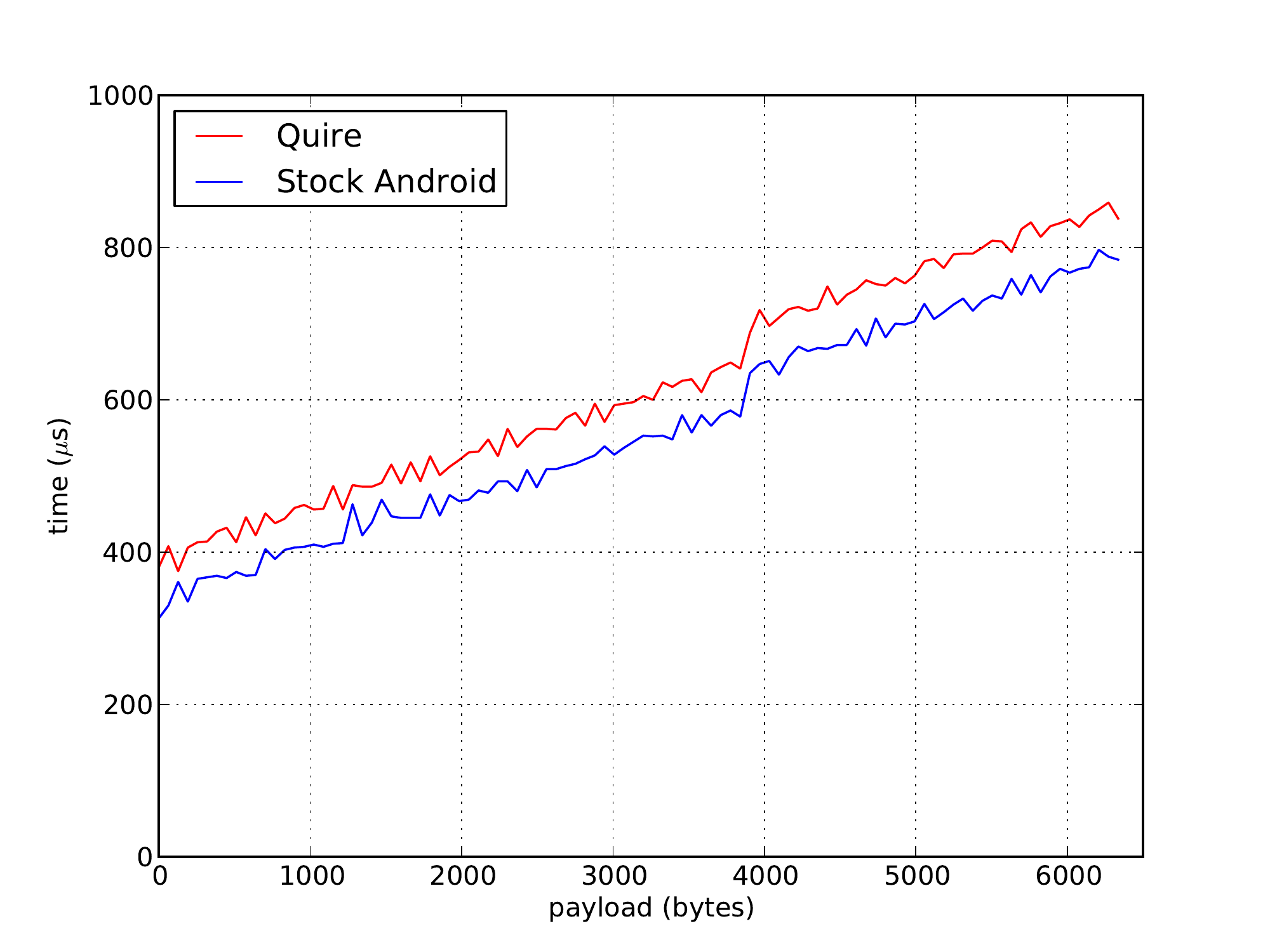}
    \caption{Roundtrip single step IPC time vs payload size.}
    \label{fig:ipc-one-step}
\end{figure}

We discarded performance timings for the first IPC call of each run
to remove any noise that could have been caused by previous activity
on the system. The results in Figure~\ref{fig:ipc-one-step} were obtained
by performing 10 runs of 100 trials each at each size point, with sizes
ranging from 0 to 6336 bytes in 64-byte increments.

These results show that the overhead of tracking the call chain for one hop
is around 70 microseconds, which is a 21\% slowdown in the worst case of doing
no-op calls.

We also measured the effect of adding a second hop into the call chain. This
was done by having two services, where the first service merely calls the
second service, which once again performs no action.

\begin{figure}[t]
    \centering
    \includegraphics[width=3in,height=2.25in]{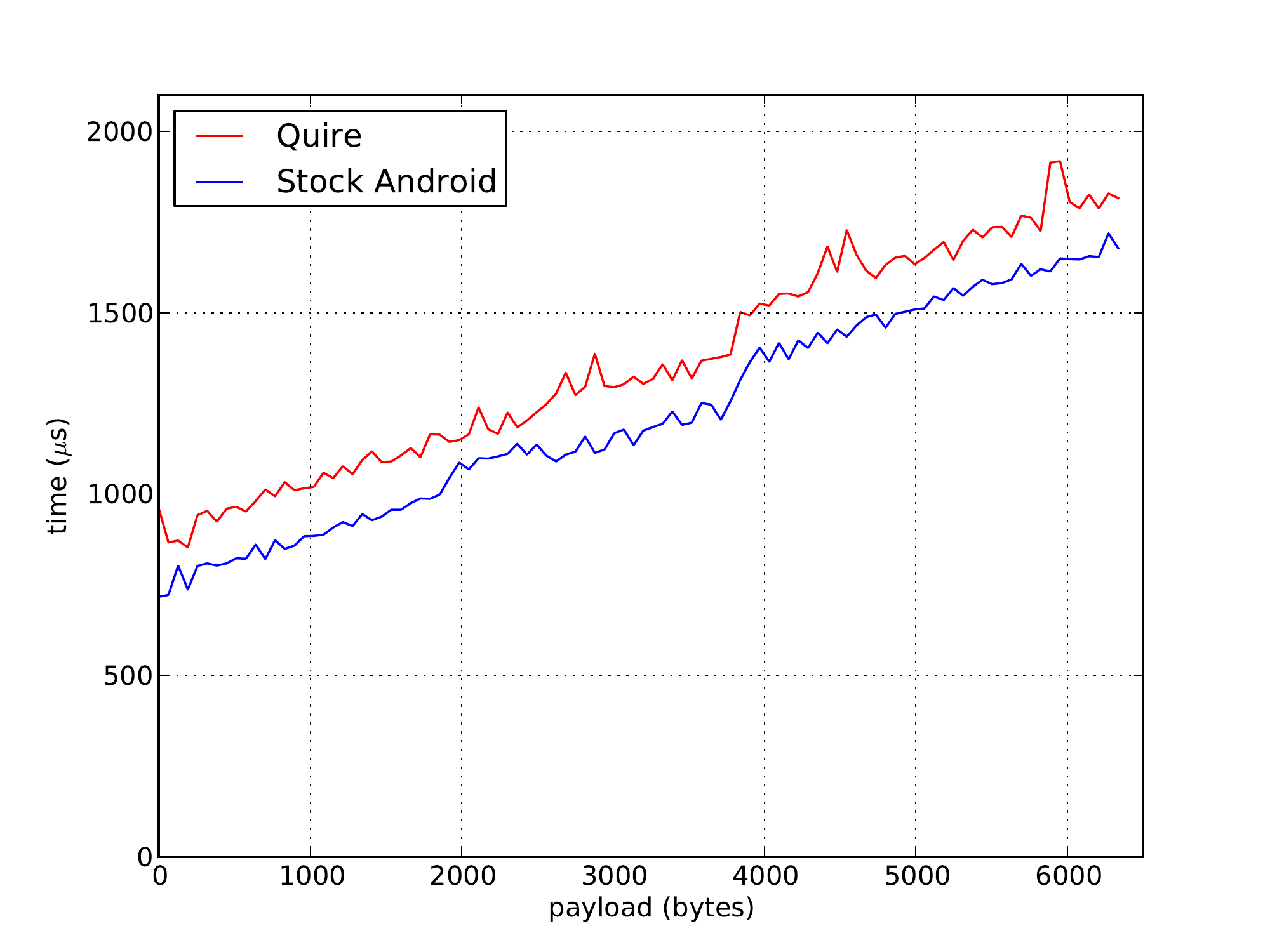}
    \caption{Roundtrip two step IPC time vs payload size.}
    \label{fig:ipc-two-steps}
\end{figure}

The results in Figure~\ref{fig:ipc-two-steps} show that the overhead of 
tracking the call chain for two hops avreages 145 microseconds, which is a 20\%
slowdown in the worst case (or, in other words, the overhead introduced by the \projname IPC
appear to be a constant factor above stock Android IPC, regardless of the call chain length). 

\subsubsection{RPC communication}

\begin{table}[h]
  \centering
  \begin{tabular}{  c | r  }
    Statement Depth & Time ($\mu$s) \\ \hline 1 & 770 \\ \hline 2 &
    1045 \\ \hline 4 & 1912 \\ \hline 8 & 4576 \\
  \end{tabular}
  \caption{IPC principal to RPC principal resolution time.}
\end{table}

The next microbenchmark we performed was determining the cost of
converting from an IPC call-chain into a serialized form that is
meaningful to a remote service.  This includes the IPC overhead in
asking the system services to perform this conversion.

We found that, even for very long statement chains, the extra cost of
this computation is a small number of milliseconds, which is
irrelevant next to the other costs associated with setting up and
maintaining a TLS network connection. From this, we conclude that
\projname RPCs introduce no meaningful overhead beyond the costs
already present in conducting RPCs over cryptographically secure
connections.

\subsection{HTTPS RPC benchmark}

To understand the impact of using \projname for calls to remote servers, we
performed some simple RPCs using both \projname and a regular HTTPS connection.
We called a simple {\em echo} service that returned a parameter
that was provided to it. This allowed us to easily measure the effect of
payload size on latency. We ran these tests on a small LAN with a single
wireless router and server plugged into this router, and using the phone's
WiFi antenna for connectivity. Each data point is the mean of 10 runs of
100 trials each, with the highest and lowest times thrown out prior to taking
the mean to remove anomalies.

\begin{figure}
    \centering
    \includegraphics[width=3in,height=2.25in]{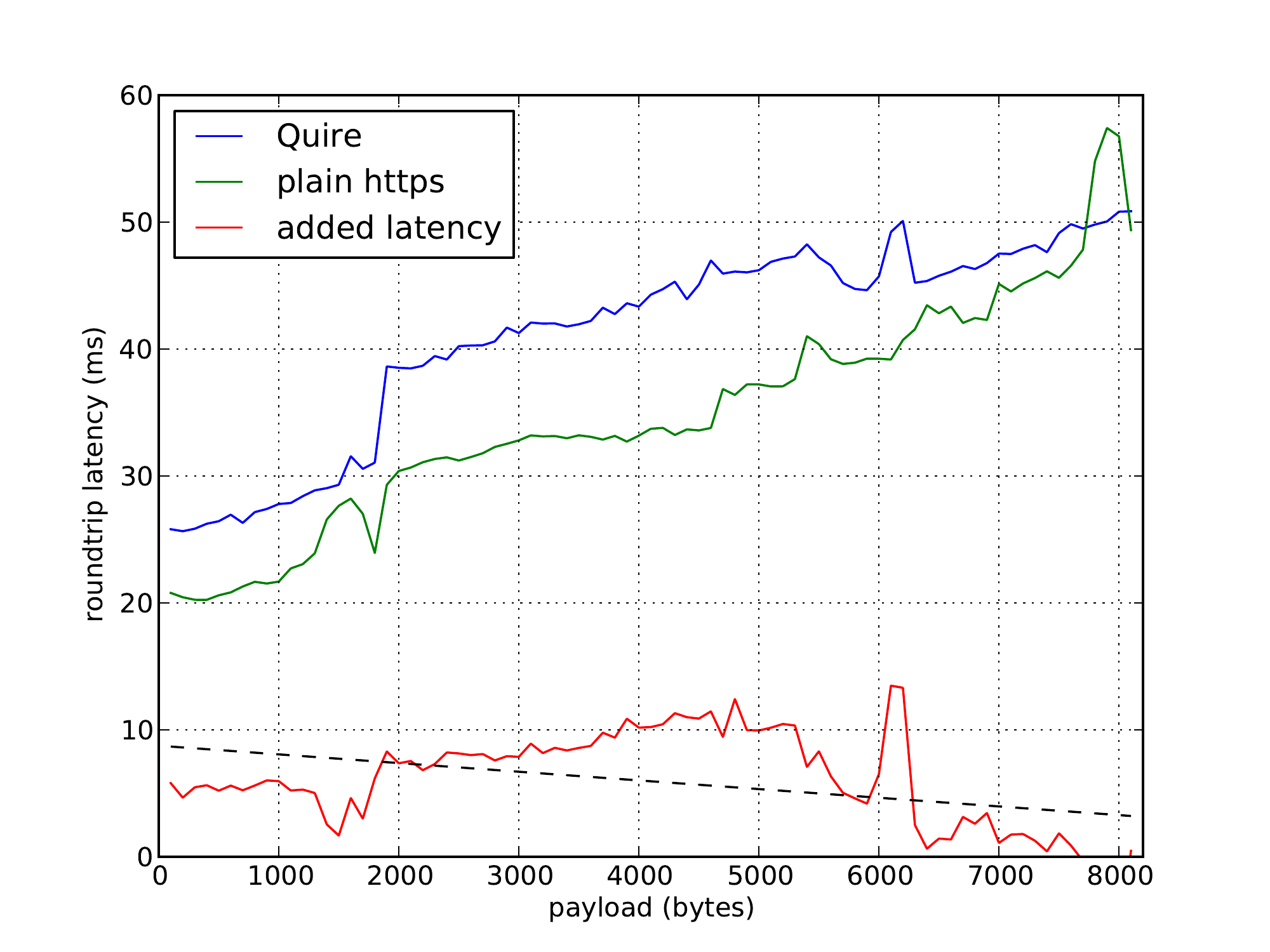}
    \caption{Network RPC latency in milliseconds.}
    \label{fig:networkrpc}
\end{figure}

The results in Figure \ref{fig:networkrpc} show that \projname adds 
an additional overhead which
averages around 6 ms, with a maximum of 13.5 ms, and getting smaller as
the payload size increases. This extra latency is small enough that 
it's irrelevant in the face of the latencies experienced across typical cellular Internet connections.
From this we can conclude that the overhead of \projname for network RPC
is practically insignificant.

\subsection{Analysis}

Our benchmarks demonstrate that adding call-chain tracking can be done
without a significant performance penalty above and beyond that of performing
standard Android IPCs. Also, the cost of creating a signed statement is low enough that
it can easily be performed for every touch event generated by the system. Finally, our
RPC benchmarks show that the addition of \projname does not cause a significant
slowdown relative to standard TLS-encrypted communications.


%% file: related.tex
\subsection{Smart phone platform security}
As mobile phone hardware and software increase in complexity the
security of the code running on a mobile devices has become a major
concern.

The Kirin system~\cite{kirin} and
Security-by-Contract~\cite{SecbyC} focus on enforcing install time
application permissions within the Android OS and .NET framework
respectively.
These approaches to mobile phone security allow a user
to protect themselves by enforcing blanket restrictions on what
applications may be installed or what installed applications may do, but do little to protect the user from
applications that collaborate to leak data or protect applications
from one another.

Saint~\cite{saint} extends the functionality of the
Kirin system to allow for runtime inspection of the full system
permission state before launching a given application.
Apex~\cite{nauman2010apex} presents another solution for the same
problem where the user is responsible for defining run-time constraints on top
of the existing Android permission system.  Both of these approaches
allow users to specify static policies to shield
themselves from malicious applications, but don't allow apps to make dynamic
policy decisions.

CRePE~\cite{conti2011crepe} presents a solution that attempts to artificially restrict an
application's permissions based on environmental constraints such as
location, noise, and time-of-day.  While CRePE considers contextual
information to apply dynamic policy decisions, it does not attempt to
address privilege escalation attacks.

\subsubsection{Dynamic taint analysis on Android}
The TaintDroid~\cite{taintdroid} and
ParanoidAndroid~\cite{paranoidAndroid} projects present dynamic
taint analysis techniques to preventing runtime attacks and data
leakage.
These projects attempt to tag objects with metadata in order to track
information flow and enable policies based on the path that data
has taken through the system.  TaintDroid's approach to information
flow control is to restrict the transmission of tainted data to a
remote server by monitoring the outbound network connections made from
the device and disallowing tainted data to flow along the outbound
channels.  The goal of \projname differs from that of taint analysis in
that \projname allows applications to protect sensitive data
at the source as opposed to the network output.

The low level approaches used to tag data also differ between the
projects. TaintDroid enforces its taint propagation semantics by
instrumenting an application's DEX bytecode to tag every variable,
pointer, and IPC message that flows through the system with a taint
value.  In contrast, \projname's approach requires only the 
IPC subsystem be modified with no reliance on instrumented code,
therefore \projname can work with applications that use native
libraries and avoid the overhead imparted by instrumenting code to
propagate taint values.

\subsubsection{Decentralized information flow control}
A branch of the information flow control space focuses on how to
provide taint tracking in the presence of mutually distrusting
applications and no centralized authority.  Meyer's and Liskov's work
on decentralized information flow control (DIFC)
systems~\cite{liskov-flowcontrol,meyers-difc} was the first attempt to
solve this problem.  Systems like DEFCon~\cite{defcon} and
Asbestos~\cite{asbestos} use DIFC mechanisms to dynamically apply
security labels and track the taint of events moving through a
distributed system.  These projects and \projname are similar in that
they both rely on process isolation and communication via message
passing channels that label data.  However, DEFCon cannot provide its
security guarantees in the presence of deep copying of data while
\projname can survive in an environment where deep copying is allowed
since \projname defines policy based on the call chain and ignores the
data contained within the messages forming the call chain.  Asbestos
avoids the deep copy problems of DEFCon by tagging data at the IPC
level.  
%
%
While Asbestos and \projname
use a similar approach to data tagging, the tags are used for very
different purposes.  Asbestos aims to prevent data leaks by enabling
an application to tag its data and disallow a recipient application
from leaking information that it received over an IPC channel while
\projname attempts to pre-emptively disallow data from being leaked by protecting
the resource itself, rather than allowing the resource to be accessed
then blocking leakage at the taint sink.

\subsection{Operating system security}
\projname is closely related to Taos~\cite{taos}.  Our design replaces
Taos's expensive digital signatures with relatively inexpensive HMAC
authenticators.  This approach was also considered as an optimization
in practical Byzantine fault tolerance (PBFT)~\cite{pbft}.  
PBFT implementation using HMAC authenticators cannot scale to large numbers of nodes because
each node requires a unique shared secret with every other node.
However, \projname can get away with using HMACs as its authentication
mechanism because each application need only register a shared secret
with a central point of authority, the operating system.  Network communication
in \projname replaces the HMACs with statements made through a cryptographically
authenticated channel.

\subsection{Trusted platform management}
Our use of a central authority for the authentication of statements
within \projname shares some similarities with projects in the trusted
platform management space.  Terra~\cite{terra} and
vTPM~\cite{perez-vtpm} both use virtual machines as the mechanism for
enabling trusted computing.  The architecture of multiple segregated
guest operating systems running on top of a virtual machine manager is
similar to the Android design of multiple segregated users running on
top of a common OS.  However, these approaches both focus on
establishing the user's trust in the environment rather than trust
between applications running within the system.

\subsection{Web security}
Many of the problems of provenance and application separation
addressed in \projname are directly related to the challenge of
enforcing the same origin policy from within the web browser.
Google's Chrome browser~\cite{barth2009security,reis2009browser}
presents one solution where origin content is segregated into distinct
processes.  Microsoft's Gazelle~\cite{gazelle} project takes this idea
a step further and builds up hardware-isolated protection domains in
order to protect principals from one another.
MashupOS~\cite{howell2007mashupos} goes even further and builds OS
level mechanisms for separating principals while still allowing for
mashups.

All of these approaches are more interested in protecting principals
from each other than in building up the communication mechanism between
principals.  \projname gets application separation for free by virtue
of Android's process model, and focuses on the expanding the capabilities of the
communication mechanism used between applications on the phone and the
outside world.

\subsection{Remote procedure calls}
For an overview of some of the challenges and threats surrounding
authenticated RPC, see Weigold et~al.~\cite{remoteclientauth}.  There
are many other systems which would allow for secure remote procedure
calls from mobile devices. Kerberos~\cite{kerberos-rfc1510} is one
solution, but it involves placing too much trust in the ticket
granting server (the phone manufacturers or network providers, in our
case).  Another potential is OAuth~\cite{oauth}, where services
delegate rights to one another, perhaps even within the phone.  This
seems unlikely to work in practice, although individual \projname
applications could have OAuth relationships with external services and
could provide services internally to other applications on the phone.


%% file: future.tex
We see \projname as a platform for conducting a variety of interesting
security research around smartphones.

\paragraph{Usable and secure UI design}
\label{sec:future:ui}
The IPC extensions \projname introduces to the Android operating
system can be used as a building block in the design and
implementation of a secure user interface. We have already demonstrated
how the system can efficiently sign every UI event, allowing for these
events to be shared and delegated safely.

Any opportunity to eliminate the need for username/password dialogs
from the experience of a smartphone user would appear to be a huge
win, particularly because it's much harder for phones to display
traditional trusted path signals, such as modifications to the chrome
of a web browser. Instead, we can leverage the low-level
client-authenticated RPC channels to achieve high-level single-sign-on
goals. Our PayBuddy application demonstrated the possibility of
building single-sign-on systems within \projname. Extending this to
work with multiple CAs or to integrate with OpenID / OAuth services
would seem to be a fruitful avenue to pursue.

\paragraph{License verification}
Google's Android team recently published an API for applications that
wish to use the Android Marketplace application to establish the
licensing validity of an installed instance of an application.  This
license verification system consists of two parts.  First the Android
Marketplace application, which facilitates the remote communication with
Google's servers in order to look up the licensing information for a
phone and secondly the License Verification Library (LVL), a bit of
third party code that facilitates communication locally with the
Marketplace app.  Immediatly after the announcement of this system, an
attack was presented~\cite{case-marketplace} in which an
attacker can disassemble and modify the function of the LVL so that it
interprets a response from the Marketplace application that indicates the application 
using the LVL is not licensed for the phone as an approval for use
rather than disapproval.  This attack could be easily prevented with the
\projname extensions to Android's IPC mechanism.

LVL would run as a separate service, with its own user-id, on the
Android phone. Any application that wishes to make use of the LVL
would query it, which would then either query the Android Marketplace
or keep a local policy cache, ultimately yielding a signed statement
in return to the caller.

\paragraph{Web browsers}
While \projname is targeted at the needs of smartphone
applications, there is a clear relationship between these and the
needs of web applications in modern browsers. Extensions to \projname could have
ramifications on how code plugins (native code or otherwise) interact
with one another and with the rest of the Web. Extensions to \projname could also
form a substrate for building a new generation of browsers with
smaller trusted computing bases, where the elements that compose a web
page are separated from one another.  This contrasts with
Chrome~\cite{reis2009browser}, where each web page
runs as a monolithic entity. Our \projname work could lead to
infrastructure similar, in some respects, to Gazelle~\cite{gazelle},
which separates the principals running in a given web page, but lacks our
proposed provenance system or sharing mechanisms.

An interesting challenge is to harmonize the differences between web
pages, which increasingly operate as applications with long-term state
and the need for additional security privileges, and applications (on
smartphones or on desktop computers), where the principle of least
privilege~\cite{saltzer75} is seemingly violated by running every
application with the full privileges of the user, whether or not this
is necessary or desirable.

%% file: conclusion.tex
In this paper we presented \projname, a set of extensions to the Android
operating system that enable applications to propagate call chain
context to downstream callees and to authenticate the origin of data that they receive indirectly.
When remote communication is needed, our RPC subsystem allows the operating
system to embed attestations about message origins and the IPC call chain
into the request. This allows remote servers to make policy decisions
based on these attestation.

We implemented the \projname design as a backwards-compatible extension to the Android
operating system that allows existing Android applications to co-exist
with applications that make use of \projname's services.

We evaluated our implementation of the \projname design by measuring
our modifications to Android's Binder IPC system with a series of
microbenchmarks. We also implemented two applications which use these
extensions to provide click fraud prevention and in-app micropayments.

Our work shows that a Taos-style system, with applications tracking call chains
and making signed statements to one another, can be implemented
efficiently on a mobile platform, enabling a variety of novel security
applications.


%% file: main.bbl
\begin{thebibliography}{10}

\bibitem{ablp91}
M.~Abadi, M.~Burrows, B.~Lampson, and G.~D. Plotkin.
\newblock A calculus for access control in distributed systems.
\newblock {\em {ACM} Transactions on Programming Languages and Systems},
  15(4):706--734, Sept. 1993.

\bibitem{barth08csrf}
A.~Barth, C.~Jackson, and J.~C. Mitchell.
\newblock Robust defenses for cross-site request forgery.
\newblock In {\em 15th ACM Conference on Computer and Communications Security
  (CCS '08)}, Alexandria, VA, Oct. 2008.

\bibitem{barth2009security}
A.~Barth, C.~Jackson, and C.~Reis.
\newblock The security architecture of the {Chromium} browser.
\newblock Technical Report,
  \url{http://www.adambarth.com/papers/2008/barth-jackson-reis.pdf}, 2008.

\bibitem{perez-vtpm}
S.~Berger, R.~C\'{a}ceres, K.~A. Goldman, R.~Perez, R.~Sailer, and L.~van
  Doorn.
\newblock {vTPM}: virtualizing the trusted platform module.
\newblock In {\em 15th Usenix Security Symposium}, Vancouver, B.C., Aug. 2006.

\bibitem{case-marketplace}
J.~Case.
\newblock Report: {Google's Android Market} license verification easily
  circumvented, will not stop pirates.
\newblock \textsf{http://www.androidpolice.com/2010/08/23/
  exclusive-report-googles-android-market-license-verification-easily-circumve%
nted-will-not-stop-pirates/}, Aug. 2010.

\bibitem{pbft}
M.~Castro and B.~Liskov.
\newblock {Practical Byzantine} fault tolerance and proactive recovery.
\newblock {\em ACM Transactions on Computer Systems (TOCS)}, 20(4):398--461,
  2002.

\bibitem{conti2011crepe}
M.~Conti, V.~T.~N. Nguyen, and B.~Crispo.
\newblock {CRePE}: Context-related policy enforcement for {Android}.
\newblock In {\em Proceedings of the Thirteen Information Security Conference
  (ISC '10)}, Boca Raton, FL, Oct. 2010.

\bibitem{SecbyC}
L.~Desmet, W.~Joosen, F.~Massacci, P.~Philippaerts, F.~Piessens, I.~Siahaan,
  and D.~Vanoverberghe.
\newblock {Security-by-contract} on the {.NET} platform.
\newblock {\em Information Security Technical Report}, 13(1):25--32, 2008.

\bibitem{taintdroid}
W.~Enck, P.~Gilbert, C.~Byung-gon, L.~P. Cox, J.~Jung, P.~McDaniel, and S.~A.
  N.
\newblock {TaintDroid}: An information-flow tracking system for realtime
  privacy monitoring on smartphones.
\newblock In {\em Proceeding of the 9th USENIX Symposium on Operating Systems
  Design and Implementation (OSDI '10)}, pages 393--408, 2010.

\bibitem{kirin}
W.~Enck, M.~Ongtang, and P.~McDaniel.
\newblock On lightweight mobile phone application certification.
\newblock In {\em 16th ACM Conference on Computer and Communications Security
  (CCS '09)}, Chicago, IL, Nov. 2009.

\bibitem{terra}
T.~Garfinkel, B.~Pfaff, J.~Chow, M.~Rosenblum, and D.~Boneh.
\newblock Terra: A virtual machine-based platform for trusted computing.
\newblock In {\em Proceedings of the 19th {S}ymposium on {O}perating {S}ystem
  {P}rinciples ({SOSP '03})}, Bolton Landing, NY, Oct. 2003.

\bibitem{oauth}
E.~Hammer-Lahav, D.~Recordon, and D.~Hardt.
\newblock {The OAuth 2.0 Protocol}.
\newblock \url{http://tools.ietf.org/html/draft-ietf-oauth-v2-10}, 2010.

\bibitem{hardy88}
N.~Hardy.
\newblock The confused deputy.
\newblock {\em ACM Operating Systems Review}, 22(4):36--38, Oct. 1988.

\bibitem{howell2007mashupos}
J.~Howell, C.~Jackson, H.~J. Wang, and X.~Fan.
\newblock {MashupOS: Operating system abstractions for client mashups}.
\newblock In {\em Proceedings of the 11th USENIX Workshop on Hot Topics in
  Operating Systems (HotOS '07)}, pages 1--7, 2007.

\bibitem{ioannidis.bellovin.ea:sub-operating}
S.~Ioannidis, S.~M. Bellovin, and J.~Smith.
\newblock Sub-operating systems: A new approach to application security.
\newblock In {\em SIGOPS European Workshop}, Sept. 2002.

\bibitem{kerberos-rfc1510}
J.~T. Kohl and C.~Neuman.
\newblock The {Kerberos} network authentication service {(V5)}.
\newblock \url{http://www.ietf.org/rfc/rfc1510.txt}, Sept. 1993.

\bibitem{defcon}
M.~Migliavacca, I.~Papagiannis, D.~M. Eyers, B.~Shand, J.~Bacon, and
  P.~Pietzuch.
\newblock {DEFCON}: high-performance event processing with information
  security.
\newblock In {\em Proceedings of the 2010 USENIX Annual Technical Conference},
  Boston, MA, June 2010.

\bibitem{jflow}
A.~C. Myers.
\newblock {JFlow}: Practical mostly-static information flow control.
\newblock In {\em Proceedings of the 26th ACM SIGPLAN-SIGACT Symposium on
  Principles of Programming Languages (POPL '99)}, pages 228--241, 1999.

\bibitem{liskov-flowcontrol}
A.~C. Myers and B.~Liskov.
\newblock A decentralized model for information flow control.
\newblock {\em ACM SIGOPS Operating Systems Review}, 31(5):129--142, 1997.

\bibitem{myers98}
A.~C. Myers and B.~Liskov.
\newblock Complete, safe information flow with decentralized labels.
\newblock In {\em Proceedings of the 1998 {IEEE} Symposium on Security and
  Privacy}, pages 186--197, Oakland, California, May 1998.

\bibitem{meyers-difc}
A.~C. Myers and B.~Liskov.
\newblock Protecting privacy using the decentralized label model.
\newblock {\em ACM Transactions on Software Engineering and Methodology
  (TOSEM)}, 9(4):410--442, 2000.

\bibitem{nauman2010apex}
M.~Nauman, S.~Khan, and X.~Zhang.
\newblock {Apex}: extending {Android} permission model and enforcement with
  user-defined runtime constraints.
\newblock In {\em Proceedings of the 5th ACM Symposium on Information, Computer
  and Communications Security}, pages 328--332, 2010.

\bibitem{saint}
M.~Ongtang, S.~McLaughlin, W.~Enck, and P.~McDaniel.
\newblock Semantically rich application-centric security in {Android}.
\newblock In {\em Proceedings of the 25th Annual Computer Security Applications
  Conference (ACSAC '09)}, Honolulu, HI, Dec. 2009.

\bibitem{paranoidAndroid}
G.~Portokalidis, P.~Homburg, K.~Anagnostakis, and H.~Bos.
\newblock Paranoid {Android}: Zero-day protection for smartphones using the
  cloud.
\newblock In {\em Annual Computer Security Applications Conference (ACSAC
  '10)}, Austin, TX, Dec. 2010.

\bibitem{reis2009browser}
C.~Reis, A.~Barth, and C.~Pizano.
\newblock Browser security: lessons from {Google Chrome}.
\newblock {\em Communications of the ACM}, 52(8):45--49, 2009.

\bibitem{saltzer75}
J.~H. Saltzer and M.~D. Schroeder.
\newblock The protection of information in computer systems.
\newblock {\em Proceedings of the {IEEE}}, 63(9):1278--1308, Sept. 1975.

\bibitem{asbestos}
S.~VanDeBogart, P.~Efstathopoulos, E.~Kohler, M.~Krohn, C.~Frey, D.~Ziegler,
  F.~Kaashoek, R.~Morris, and D.~Mazi\`{e}res.
\newblock Labels and event processes in the {Asbestos} operating system.
\newblock {\em ACM Transactions on Computer Systems (TOCS)}, 25(4), Dec. 2007.

\bibitem{wallach98}
D.~S. Wallach and E.~W. Felten.
\newblock Understanding {J}ava stack inspection.
\newblock In {\em Proceedings of the 1998 {IEEE} Symposium on Security and
  Privacy}, pages 52--63, Oakland, California, May 1998.

\bibitem{wallach-tosem2000}
D.~S. Wallach, E.~W. Felten, and A.~W. Appel.
\newblock The security architecture formerly known as stack inspection: A
  security mechanism for language-based systems.
\newblock {\em ACM Transactions on Software Engineering and Methodology},
  9(4):341--378, Oct. 2000.

\bibitem{gazelle}
H.~J. Wang, C.~Grier, A.~Moshchuk, S.~T. King, P.~Choudhury, and H.~Venter.
\newblock The multi-principal {OS} construction of the {Gazelle} web browser.
\newblock In {\em Proceedings of the 18th USENIX Security Symposium}, 2009.

\bibitem{remoteclientauth}
T.~Weigold, T.~Kramp, and M.~Baentsch.
\newblock Remote client authentication.
\newblock {\em IEEE Security \& Privacy}, 6(4):36--43, July 2008.

\bibitem{taos}
E.~Wobber, M.~Abadi, M.~Burrows, and B.~Lampson.
\newblock Authentication in the {Taos} operating system.
\newblock {\em ACM Transactions on Computer Systems (TOCS)}, 12(1):3--32, 1994.

\bibitem{dstar2008}
N.~Zeldovich, S.~Boyd-Wickizer, and D.~Mazi\`{e}res.
\newblock Securing distributed systems with information flow control.
\newblock In {\em Proceedings of the 5th Symposium on Networked Systems Design
  and Implementation (NSDI '08)}, San Francisco, CA, Apr. 2008.

\end{thebibliography}
